\documentclass[reprint,amsmath,amssymb,aps,nofootinbib]{revtex4-1}

\usepackage{graphicx}
\usepackage{dcolumn}
\usepackage{bm}

\usepackage{color}
\usepackage{xcolor}

\usepackage{mathrsfs}
\usepackage{amsfonts}
\usepackage{varioref}
\usepackage{csquotes}
\usepackage{hyphenat}
\usepackage{float}
\usepackage{multirow}

\RequirePackage[colorlinks,citecolor=blue,urlcolor=magenta,linkcolor=blue]{hyperref}

\labelformat{chapter}{Chapter (#1)}
\labelformat{section}{Section (#1)}
\labelformat{subsection}{Section (#1)}
\labelformat{subsubsection}{Section (#1)}
\labelformat{subsubsubsection}{Section (#1)}
\labelformat{equation}{Eq.~(#1)}
\labelformat{figure}{Fig.~(#1)}
\labelformat{subfigure}{Fig.~(\thefigure#1)}
\labelformat{table}{Tab.~(#1)}
\labelformat{appendix}{Appendix #1}
\begin{document}

\title{Pure Gauss-Bonnet NUT black hole with and without non-central singularity}

\author{Sajal Mukherjee}
\email{sajal@iucaa.in}

\author{Naresh Dadhich}
\email{nkd@iucaa.in}

\affiliation{Inter-University Centre for Astronomy and Astrophysics, Post Bag 4, Pune-411007, India}

\begin{abstract}
\noindent
It is known that NUT solution has many interesting features and pathologies like being non-singular and having closed timelike curves. It turns out that in higher dimensions horizon topology cannot be spherical but it has instead to be product of $2$-spheres so as to retain radial symmetry of spacetime. In this letter we wish to present a new solution of pure Gauss-Bonnet $\Lambda$-vacuum equation describing a black hole with NUT charge. It has three interesting cases: (a) black hole with both event and cosmological horizons with singularity being hidden behind the former, (b) a regular spacetime free of both horizon and singularity, and (c) black hole with event horizon without singularity and cosmological horizon. Singularity here is always non-centric at $r \neq 0$.

\end{abstract}
\maketitle
%
NUT spacetime is the most enigmatic solution of the Einstein vacuum equation. It is the most general radially symmetric solution when asymptotic flatness condition is lifted off. Its physical interpretation and understanding has very rich literature, see for instance \cite{Newman:1963yy,Misner:1963fr,Bonnor:NUT, LyndenBell:1996xj,Turakulov:2001bm,Turakulov:2001jc,Mukherjee:2018dmm,Jefremov:2016dpi}. It is non-singular and asymptotically non-flat black hole solution with Kretschmann curvature scalar being  finite at $r=0$. It, however, admits closed timelike curves \cite{Astefanesei:2004kn}.

When we go to higher dimensions, a spectrum of topology opens up for black hole horizon from spherical to product of spheres,  for instance in six dimension, it could be $S^{(4)}$ or $S^{(2)} \times S^{(2)}$ or $S^{(1)} \times S^{(3)}$ \cite{Hervik:2019gly,Hervik:2020zvn}. Interestingly it turns out that all higher dimensional NUT solutions do not have spherical topology but instead have product of $2$-spheres \cite{Awad:2000gg,Awad:2005ff,Dehghani:2005zm,Mann:2005mb,Chen:2006xh,Flores-Alfonso:2018jra}. This is because spherical topology in dimension, $D-2 > 2$ is not compatible with radial symmetry, which is a primary property of NUT spacetime \cite{Mukherjee:2021erg}. Hence in higher dimensions NUT black hole would always have topology of product of $S^{(2)}$ spheres.

When we go over from general relativity (GR) to Gauss-Bonnet (GB) gravity in higher dimensions \cite{Lovelock:1971yv,Boulware:1985wk,Zwiebach:1985uq}, there comes in another new feature due to occurrence of discriminant in the solution, which has to be non-negative for reality of spacetime. Discriminant also occurs in the denominator in curvatures indicating a non-centric singularity \cite{Pons:2014oya,Dadhich:2015nua}. Further a positive cosmological constant is also required for validity of solution for large $r$. Thus a parameter window in terms of mass and $\Lambda>0$ is defined for the solution to be real as well as free of non-centric singularity. Note that the occurrence of non-centric singularity is due to product topology which is present even when absence of NUT charge \cite{Pons:2014oya}. The NUT charge brings in its own complexity and makes the analysis more involved.

We have found an exact solution \cite{Mukherjee:2021erg} of pure GB $\Lambda$-vacuum equation describing with a NUT charge. Now the discriminant turns much more complex due to presence of the NUT parameter and its clubbing with mass and $\Lambda$. Again its vanishing gives non-centric singularity. There is however no central singularity due to the presence of NUT parameter. In this letter we wish to present three interesting cases of this solution, which are only physically viable. They are as follows: (a) black hole with both event and cosmological horizons with singularity being hidden behind the former, (b) a regular spacetime free of both horizon and singularity, and (c) black hole with event horizon  without singularity and cosmological horizon. We now go over to the discussion of  newly found pure GB NUT solution.

By choosing the topology of product of two spheres, $S^{(2)} \times S^{(2)}$, we have obtained a new solution \cite{Mukherjee:2021erg} of pure GB $\Lambda$-vacuum equation for a black hole with NUT charge and it is given by the metric,
\begin{eqnarray}
ds^2=-\dfrac{\Delta}{\rho^2}\Bigl\{dt + P_1 d\phi_1 +P_2d\phi_2 \Bigr\}^2+\dfrac{\rho^2}{\Delta}dr^2+\dfrac{\rho^2}{3} \Bigl\{d\theta_1^2+\nonumber \\
\sin^2\theta_1 d\phi^2_1+d\theta_2^2+\sin^2\theta_2 d\phi^2_2 \Bigr\}, \nonumber \\
\label{eq:metric_S2_S2_SNUT}
\end{eqnarray}
where $l$ is the NUT parameter, $\rho^2=r^2+l^2$, $P_1={2 l \cos\theta_1/3}$, $P_2={2 l \cos\theta_2/3}$ and $\Delta=\rho^2 \left(1-f(r)\right)$. The function $f(r)$ reads as
%
\begin{equation}
f(r)=\dfrac{1}{l^2-r^2}\Bigl[{2 l^2-\Big\{h(r)\Big\}^{1/2}}\Bigr],
\label{eq:f(r)}
\end{equation}
where
\begin{eqnarray}
h(r)=\Bigl[\dfrac{\Lambda}{60} r^8+\dfrac{\Lambda}{15}l^2 r^6+\dfrac{1}{6}\left(\lambda -12\right)r^4+M r^3\nonumber \\
-\dfrac{l^2 r^2}{3}\left(\lambda -24\right)-M l^2 r+\dfrac{l^4}{12}\left(\lambda -24\right)\Bigr].
\label{C_1_0_C_2_0_NEW}
\end{eqnarray}
%
We have defined the dimensionless parameter $\lambda = \Lambda l^4$. Clearly the discriminant $h(r)$ has to be non-negative for reality of the metric function $f(r)$. Further it turns out that the curvature scalar diverges at $h(r)=0$ as it occurs in the denominator and it represents a non-centric singularity at $r \neq 0$.

On the other hand horizon is given by positive roots of $h(r)-(r^2+l^2)^2=0$ which factors as $(r^2-l^2)\mathcal{X}(r) = 0$. Note that when $l^2=r^2$ both numerator and denominator vanish and the limit turns out to be finite non-zero. Hence $r^2 \neq l^2$ and  horizon would then be given by $\mathcal{X}(r) = 0$ which reads as
\begin{eqnarray}
\mathcal{X}(r)=r^6 \Lambda+5 \Lambda l^2 r^4+15 r^2(\lambda -12)+60 M r- \nonumber \\
5 l^2(\lambda -36)=0.
\label{eq:h_tilder_h}
\end{eqnarray}
We have therefore to find parameter windows for which either $h(r) > 0$ or $h(r) = 0$ occurs inside the event horizon so that singularity is not naked. It turns out that there exist three distinct such parameter windows which we now take up.

In view of \ref{C_1_0_C_2_0_NEW} and \ref{eq:h_tilder_h} there arise only three viable cases depending on values of the parameter $\lambda = \Lambda l^4$ and they are: (a) $\lambda <12$, there occur both event and cosmological horizons and a non-naked singularity hidden behind the former. (b) $24 < \lambda < 36$, there are neither horizons nor singularity, a regular spacetime without black hole. (c) $\lambda >36$, there occurs only event horizon without singularity and cosmological horizon. These three cases we discuss are all viable as we demonstrate by non-existence of a naked singularity. A spacetime with naked singularity is only non-viable, else all are viable, however, they may or may not be BHs. Here we have excluded the range $12 \leq \lambda \leq 24$ because for this singularity always occurs and it remains naked as there is no horizon to cover it. Hence this parameter window is prohibited. Of course, we would like to seek the cases that have either no singularity, or if it occurs, it is hidden behind event horizon. The positive roots of \ref{C_1_0_C_2_0_NEW} and \ref{eq:h_tilder_h} represent singularity and horizon respectively. What is therefore desired is that the former should have no or at best one positive root which could be covered by event horizon while the latter could have one or two or none positive roots respectively representing event, event and cosmological and no horizon. When there occurs no horizon, there must not occur singularity either else it would be naked.

We now exhibit these three case with specific examples. In all Figs below, black and red curve respectively refer to $h(r)$ and $\mathcal{X}(r)$. Besides, these expressions are scaled by $r^4$ for better representation of our results.
%

\begin{figure}[htp]
\includegraphics[scale=.3]{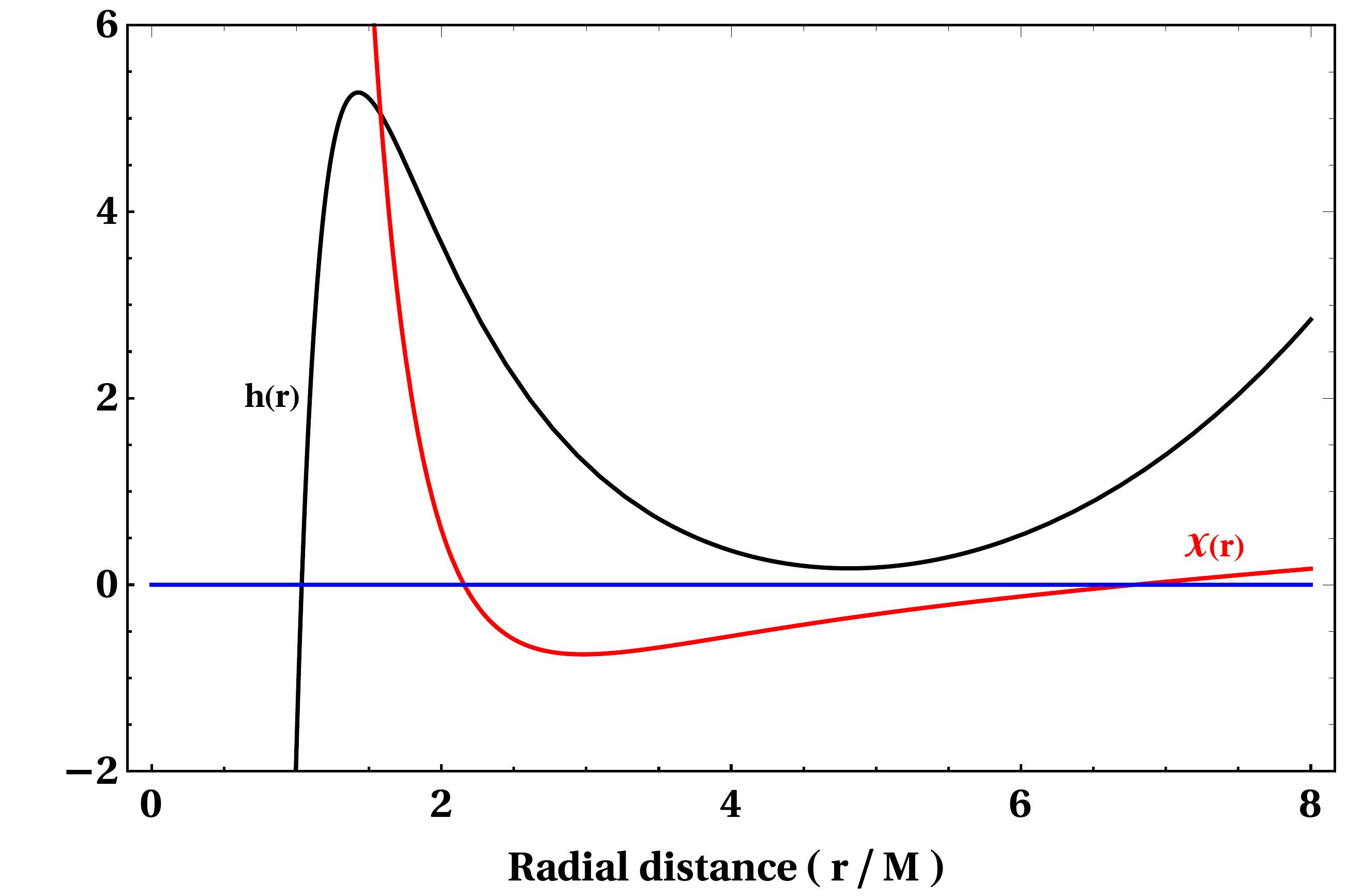}
\caption{In this case, the singularity is hidden behind the event horizon and there also occurs cosmological horizon.
}
\label{fig:Figure_01}
\end{figure}

(a) $\lambda <12$~:--- In this case, \ref{C_1_0_C_2_0_NEW} will always have at least one positive root while \ref{eq:h_tilder_h} can have utmost two. Therefore one singularity is unavoidable and needs to be covered by event horizon. There could however occur two horizons, event and cosmological. So we have the case of singularity being hidden behind the event horizon and there also occurs cosmological horizon as  illustrated in \ref{fig:Figure_01} for $\Lambda M^4=0.05$ and $l=1.805M$.
Given that we have to select a parameter space where both singularity and horizons exist, the limit on $l$ is given for $\Lambda M^4=0.05$ as:
\begin{equation}
0<1.79 M < l < 1.82M<\left\{12/\Lambda\right\}^{1/4} \simeq 3.94M.
\end{equation}
Clearly the NUT parameter $l$ has a fairly narrow window.

\begin{figure}[htp]
\includegraphics[scale=.3]{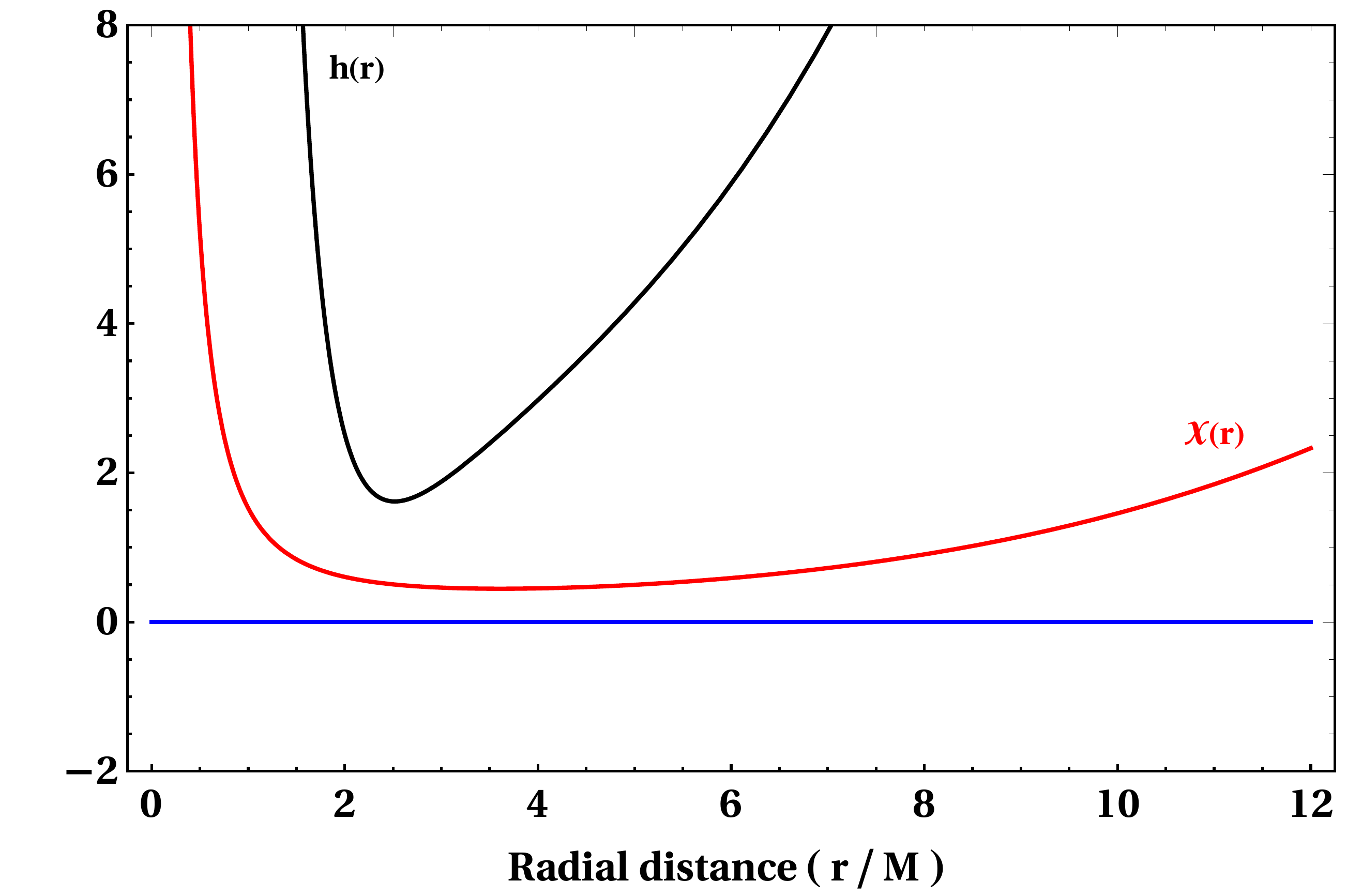}
\caption{There occurs neither singularity nor horizon.}
\label{fig:Figure_02}
\end{figure}

(b) $24<\lambda <36$~:--- Note that for the range  $12\leq \lambda \leq 36$, \ref{eq:h_tilder_h} has no positive root indicating there exists no horizon. In the range $24<\lambda <36$, \ref{C_1_0_C_2_0_NEW} can have zero or two positive roots indicating possibility of singularity free spacetime. In \ref{fig:Figure_02}, we provide an example of a NUT compact object having neither horizon nor singularity for $\Lambda M^4=0.05$ and $l=4.8M$.
Moreover, as \ref{C_1_0_C_2_0_NEW} can also have two positive roots and to avoid them may further restrict range for the NUT parameter $l$. In particular with $\Lambda M^4=0.05$, the bounds on $l$ can be expressed as
\begin{equation}
\left(\dfrac{24}{\Lambda}\right)^{1/4} \simeq 4.68M <4.72 M<l<\left(\dfrac{36}{\Lambda}\right)^{1/4} \simeq 5.18M,
\end{equation}
which is severely constrained.

\begin{figure}[htp]
\includegraphics[scale=.3]{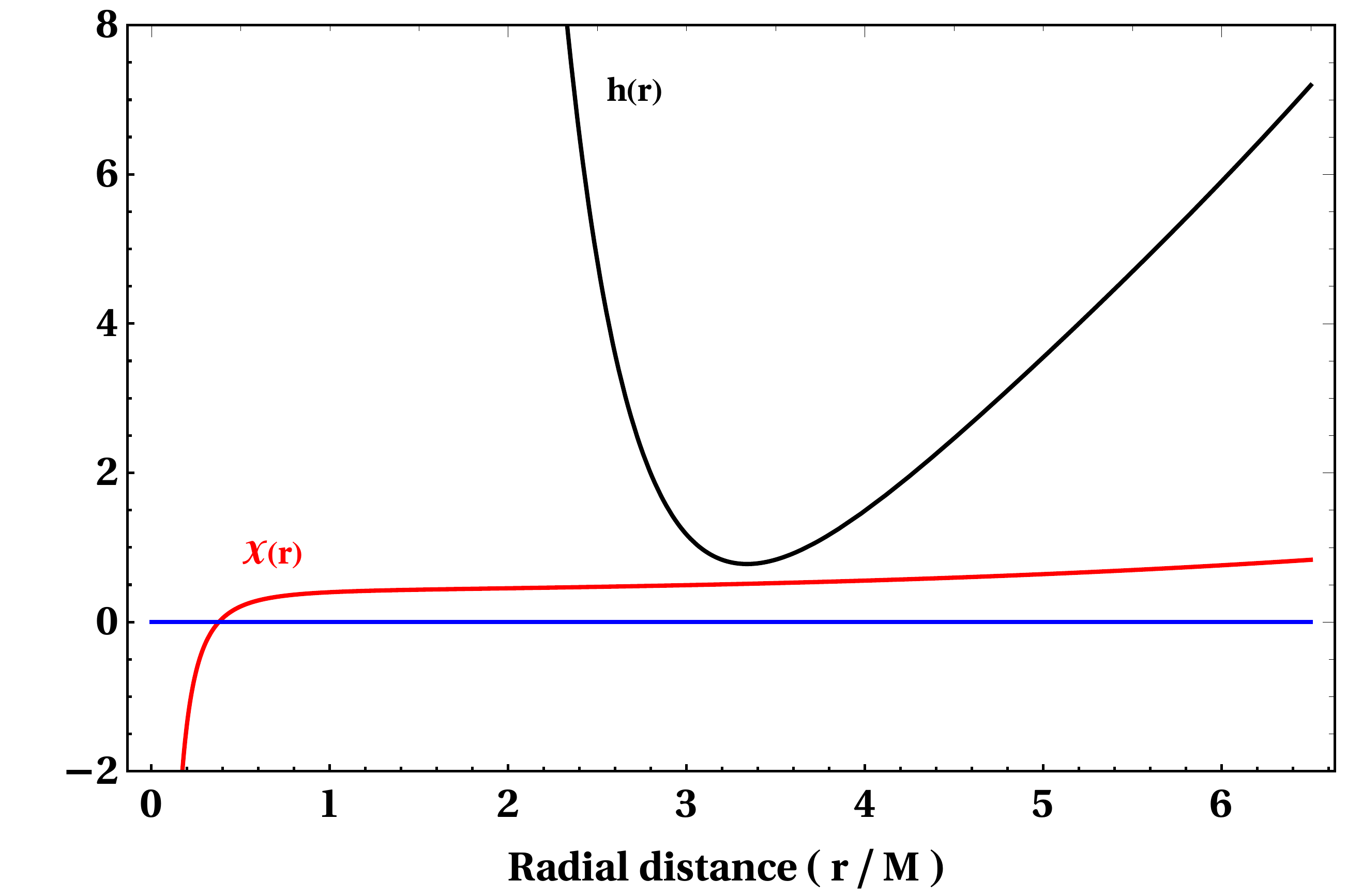}
\caption{A black hole with only event horizon and free of cosmological horizon as well as singularity. }
\label{fig:Figure_03}
\end{figure}

(c) $\lambda >36$~:---
In this case \ref{C_1_0_C_2_0_NEW} can have zero or two positive roots while \ref{eq:h_tilder_h} will always have one positive root. This means horizon (event) always occurs while singularity may or may not. In \ref{fig:Figure_03} we have taken $\Lambda M^4=0.05$, and $l=5.2M$ such that $\Lambda l^4 >36$. This presents a case of black hole having event horizon but no cosmological horizon.
Also note that for $\lambda >36$, $h(r)=0$ can also have two positive roots. To avoid them, NUT charge gets further constrained, and for $\Lambda M^4=0.05$, it becomes
\begin{equation}
(36/\Lambda)^{1/4} \simeq 5.18M<l<5.41M
\end{equation}
which further squeezes the window for $l$.

We have obtained an exact solution of pure GB $\Lambda$-vacuum solution with a NUT parameter. The physical viability of spacetime does seem to severely restrict window for the NUT parameter. It should however be remembered that NUT is a gravomagnetic charge \cite{LyndenBell:1996xj,Turakulov:2001bm,Turakulov:2001jc} for which we have no idea of its physically viable value.

The characteristic feature of pure GB solutions \cite{Pons:2014oya,Dadhich:2015nua} with product topology is the occurrence of discriminant, $h(r)$ in the solution and it has to be non-negative, and further its vanishing represents a non-central curvature singularity at finite $r\neq0$. To avoid singularity it is required that $h(r) > 0$. Even if it occurs it should not be naked and hence should be hidden behind an event horizon. First and foremost it requires presence of positive $\Lambda$ else spacetime has no large $r$ viability. That is, occurrence of non-central singularity and mandatory presence of $\Lambda >0$ are purely pure GB properties \cite{Pons:2014oya,Dadhich:2015nua} even in absence of NUT charge. In the present case NUT parameter brings in its own further complexity and subtlety. Note that NUT spacetime can have only non-central singularity, which is solely inflected by GB, as it is always free of central singularity at $r=0$.

The above discussed cases seem to be the only three viable cases of the pure-GB vacuum solution. They describe the following three situations: (a) For $\lambda  < 12$, it is a black hole spacetime with singularity being covered by event horizon and there also exists a cosmological horizon. This is a usual black hole spacetime with both event and cosmological horizons and singularity hidden behind the former. (b) For $24<\lambda <36$, it is a spacetime which is without any horizon and singularity. It is regular everywhere and yet describes a compact object with NUT charge. It is interesting to note that the parameter window $12 \leq \lambda \leq 24$ is completely forbidden because in this case a naked singularity is unavoidable. (c) For $\lambda >36$, there occurs only the event horizon without singularity and cosmological horizon.

One may however ask a question, how is it that $\Lambda$ is present yet there is no cosmological horizon. For that we expand the horizon equation, $f(r)=1$ for large $r$ and find that it goes as follows:
\begin{equation}
r^6 \Lambda+5 \Lambda l^2 r^4+15 r^2(\lambda -12)=0.
\end{equation}
Clearly it has no positive root for $\lambda >12$ and hence indicating absence of cosmological horizon whenever $\lambda >12$.

On the other hand we also have the case (b) for which even event horizon doesn't exist. To understand that we take the other end limit for small $r$, which goes as
\begin{equation}
15 r^2(\lambda -12)+60 M r-5 l^2(\lambda -36)=0.
\label{eq:horizon}
\end{equation}
It would always have a positive root indicating event horizon for $\lambda >36$ while no positive roots for $12< \lambda <36$. That is why spacetime is free of horizons in the latter window.

The picture that emerges is as follow: For $\lambda <12$, we have the expected situation of a NUT black hole with both event and cosmological horizons and singularity being covered by the former. This is the most viable and acceptable case (a). As $\lambda$ increases horizons disappear until $\lambda >36$. The intermediate range, $12 \leq \lambda \leq 24$ is completely prohibited because naked singularity is inescapable while for $24 < \lambda < 36$ presents an interesting possibility of spacetime which has neither singularity nor horizons, case (b). It is very remarkable that it  is a spacetime which is regular everywhere and yet it is a vacuum solution without horizon with positive cosmological constant. Its physical character needs to be further probed and understood. For $\lambda > 36$, event, but not cosmological, horizon reappears and singularity could as well be avoided.

Physically, we do not quite understand how and why the above picture happens. This is the question for further study and engagement. The critical parameter that determines horizon and singularity structure of this spacetime is the dimensionless parameter $\lambda = \Lambda l^4$ which represents clubbing of the two parameters having opposite effects. The NUT parameter $l$ contributes to gravitational attraction in consonance with mass, while $\Lambda > 0$ has repulsive contribution.

One of us (S.M.) is thankful to the Department of Science and Technology (DST), Government of India, for financial support.
\bibliography{References}
\bibliographystyle{./utphys1}
\end{document}